\documentclass[anon]{l4dc2025}

\title[One controller to rule them all]{One controller to rule them all}

\usepackage{graphicx}
\usepackage{amsmath}
\usepackage{amsfonts}
\usepackage{caption}
\usepackage{makecell}
\usepackage{tikz}
\usepackage{standalone}
\usepackage{algorithm2e}
\usepackage{soul}
\usepackage{wrapfig}

\usepackage{footmisc}

\author{%
 \Name{Riccardo Busetto} \Email{riccardo.busetto@supsi.ch}\\
 \addr IDSIA Dalle Molle Institute for Artificial Intelligence USI-SUPSI, Lugano-Viganello, Switzerland
 \AND
 \Name{Valentina Breschi} \Email{v.breschi@tue.nl}\\
 \addr Dept. of Electrical Engineering, Eindhoven University of Technology, Eindhoven, The Netherlands
 \AND
 \Name{Marco Forgione} \Email{marco.forgione@supsi.ch}\\
 \addr IDSIA Dalle Molle Institute for Artificial Intelligence USI-SUPSI, Lugano-Viganello, Switzerland
 \AND
 \Name{Dario Piga} \Email{dario.piga@supsi.ch}\\
 \addr IDSIA Dalle Molle Institute for Artificial Intelligence USI-SUPSI, Lugano-Viganello, Switzerland
 \AND
 \Name{Simone Formentin} \Email{simone.formentin@polimi.it}\\
 \addr Dip. di Elettronica, Informazione e Bioingegneria, Politecnico di Milano, Milano, Italy
}

\begin{document}

\maketitle

\begin{abstract}
Imagine having a system to control and only know that it belongs to a certain class of dynamical systems. Would it not be amazing to simply plug in a controller and have it work as intended? With the rise of \emph{in-context learning} and powerful architectures like Transformers, this might be possible, and we want to show it. In this work, within the model reference framework, we hence propose the first in-context learning-based approach to design a unique \emph{contextual controller} for an entire class of dynamical systems rather than focusing on just a single instance. Our promising numerical results show the possible advantages of the proposed paradigm, paving the way for a shift from the \textquotedblleft one-system-one-controller\textquotedblright \ control design paradigm to a new \textquotedblleft one-class-one-controller\textquotedblright \ logic.    
\end{abstract}


\section{Introduction}
Regulating a system whose dynamics is partially or totally unknown requires data to either identify a model for the system \cite{ljung1998system} and then design a model-based controller or to directly design the controller from data (see \cite{hou2013model} for a discussion on the difference between direct and indirect data-driven control strategies). Nevertheless, existing direct and indirect control approaches generally rely on the assumption that data for control design are gathered from the plant to be regulated. Using the machine learning terminology, this is equivalent to assuming there is no domain shift \cite{farahani2021brief} between the training and the test set, which is nonetheless often unrealistic in practice. Indeed, in most real-world scenarios, data distribution shifts might occur due to several factors, e.g., changes in the environment, in the data acquisition process, or due to intrinsic differences among systems that are nominally the same. 

From the evidence of this distribution shifts in practice comes the need for tools like \emph{transfer learning} and \emph{meta learning} \cite{li2018learning, khoee2024domain} that allow one to leverage data from different, yet somehow similar, distributions for learning. Among these approaches, \textit{in-context learning} (ICL) \cite{brown2020language,dong2022survey} is a paradigm where a (sufficiently) powerful model is pre-trained with data from similar systems or tasks to immediately 
and accurately undertake a new task in a new system by relying upon a few data from the new instance of the problem \textit{without further tuning}. This aspect differentiates ICL from other meta-learning approaches that most often require further adaptation \cite{finn2017model} or fine-tuning \cite{bansal2022meta}, while being particularly appealing for control design as it would allow to undertake the design effort once and then have a running controller for an entire class of systems. In light of this intuition, ICL has recently caught the attention of the control community, with the seminal works in \cite{forgione2023system} and \cite{busetto2024context} showing that ICL could be a viable (and advantageous) solution for system identification and state estimation. Both these approaches rely on Transformers, whose \emph{attention mechanism} \cite{vaswani2017attention} has already been proven to be a powerful enough architectural choice to carry out ICL tasks in several fields, e.g., natural language processing \cite{akyurek2024context}, computer vision \cite{wang2023context} and robotics \cite{zhu2024incoro}, and whose theoretical viability to tackle state estimation is analyzed in \cite{goel2024can}.

Following the footsteps of \cite{forgione2023system,busetto2024context}, we here exploit the ICL paradigm to design a \emph{model reference}, \textit{contextual controller} that can regulate \textit{all} the systems belonging to a given class. Given the challenging task we aim to pursue, we propose a training procedure that relies on \emph{curriculum learning} to gradually increase the complexity of the learning task over training, given the success stories of its use for Transformer's training (see, e.g., \cite{xu2020curriculum,narvekar2020curriculum}). Due to our architectural choices and training strategies, learning such a powerful tool is demanding in terms of time and computational resources. However, training can be conducted offline, resulting in a unique controller ready to be deployed for an entire class of systems. As shown numerically in this paper, our approach $(i)$ allows zero-shot regulation of any system belonging to the class, $(ii)$ with no need for hyperparameters selection apart from the (rather general) structural choices for the controller, leading to $(iii)$ closed-loop performance comparable to the ones of optimal control approaches.

Though seminal, this paper represents a first step towards a shift in the control design paradigm, drifting from the practical limitation of the classical \textquotedblleft one-system-one-controller\textquotedblright \ logic, leveraging ICL to design a unique controller to regulate multiple systems.

The paper is structured as follows. The \textit{contextual control} problem is introduced in Section~\ref{sec:problem_formulation} and formulated in Section~\ref{sec:design}. Details on the contextual controller structure and training are then provided in Section~\ref{sec:training}. Its effectiveness is then assessed numerically in Section~\ref{sec:example}, with the paper ending with concluding remarks and directions for future work.
\vspace{-.15cm}

\section{Setting and goal}\label{sec:problem_formulation}
Consider a discrete-time, nonlinear system $S$, 
described by the 
\emph{unknown} state-space model
\vspace{-.15cm}
\begin{equation}\label{eq:system}
    \begin{aligned}
        x_{k+1}&=f(x_k,u_k)+w_k,\\
        y_{k}&=g(x_k)+v_k, \vspace{-.15cm}
    \end{aligned}
\end{equation}
where $x_k \in \mathbb{R}^{n_x}$ is the (inaccessible) state of the system at time $k \in \mathbb{N}$, $u_k \in \mathbb{R}^{n_u}$ and $y_k \in \mathbb{R}^{n_y}$ are the associated controllable input and measured output, while $w_k \in \mathbb{R}^{n_x}$ and $v_k \in \mathbb{R}^{n_y}$ are the realizations at time $k$ of the process and measurement noise corrupting the system, here assumed white, zero-mean, uncorrelated stochastic processes. 

Given a user-defined reference signal $r_k \in \mathbb{R}^{n_y}$ to be tracked by the output of $S$ and the information (or \emph{context}) $\mathcal{I}_k \in \mathbb{R}^{n_\mathcal{I}}$ available up to time $k$, i.e.,\vspace{-.15cm}
\begin{equation}\label{eq:information_vector}
    \mathcal{I}_k=\{e_\kappa,u_{\kappa-1}\}_{\kappa=0}^{k},\vspace{-.15cm}
\end{equation}
where $e_\kappa=e_\kappa-y_\kappa$ is the tracking error at time $\kappa \in \{0,1,\ldots,k\}$ and $u_{-1}$ is given, our objective is to design a feedback controller $\mathcal{C}_\phi: \mathbb{R}^{n_\mathcal{I}} \rightarrow \mathbb{R}^{n_u}$ such that the control sequence\vspace{-.15cm}
\begin{equation}\label{eq:controller}
u_{k}=\mathcal{C}_\phi(\mathcal{I}_k),~~k=0,1,2,\ldots\vspace{-.15cm}
\end{equation}
allows $S$ to attain a \emph{desired tracking behavior} in closed-loop. Such a target is here dictated by a pre-fixed \emph{reference model} $\mathcal{M}$, whose dynamics is:\vspace{-.15cm}
\begin{equation}\label{eq:reference_model}
    \begin{aligned}
        x_{k+1}^{M}&=f^{M}(x_{k}^{M},r_k),\\
        y_k^{d}&=g^{M}(x_{k}^{M},r_k),\vspace{-.15cm}
    \end{aligned}
\end{equation}
where $y_k^{d} \in \mathbb{R}^{n_y}$ is the output behavior we aim to mimic. Therefore, alike \cite{campi2006direct}, our goal is to solve a model reference control problem.

Assume that $f: \mathbb{R}^{n_x} \times \mathbb{R}^{n_u} \rightarrow \mathbb{R}^{n_x}$ and $g: \mathbb{R}^{n_x} \rightarrow \mathbb{R}^{n_y}$ in \eqref{eq:system} are \emph{unknown}, i.e., we do not know the \textquotedblleft true\textquotedblright \ model of $S$, and thus we cannot directly use a model-based approach to attain our design goal. Nonetheless, we have access (although, possibly, not in their explicit forms) to the \emph{prior distributions} for $(i)$ the class $\mathcal{S}$ of dynamical systems to which $S$ belongs (i.e., $p(\mathcal{S})$), $(ii)$ the references the user might be interested in tracking (namely, $p(\mathcal{R})$), and $(iii)$ the initial states and inputs ($p(\mathcal{O})$). Thanks to this information, we can generate a potentially infinite-dimensional \emph{meta-dataset} 
\begin{equation}\label{eq:meta_data}
    \mathcal{D}=\left\{S^{(i)},r_{[0,N-1]}^{(i)},\left(x_{0}^{(i)},u_{-1}^{(i)}\right)\right\}_{i=1}^{b},
\end{equation}
where $S^{(i)} \in \mathcal{S}$ is the $i$-th random system drawn from $p(\mathcal{S})$, and $x_{0}^{(i)}$ and $u_{-1}^{(i)}$ are the associated initial state and input, while $r_{[0,N-1]}^{(i)}$ compactly denotes the $i$-th reference sequence of length $N$ instantiated at random from the known set point distribution.

By using a Transformer architecture to characterize the information/input map in \eqref{eq:controller}, our goal in this work is to use this meta-dataset to solve the model-reference design problem \emph{directly} (i.e., bypassing any intermediate modeling step), hence harnessing for the first time the \emph{in-context learning} capabilities of Transformers directly for control.

\section{In-context controller design}\label{sec:design}
Toward the formulation of an \emph{in-context control design} problem, let us consider any finite horizon tracking task drawn from $p(\mathcal{R})$, with $r_{[0,N-1]}$ being the associated reference sequence. Let us further assume that, for any such set point, there exists a finite, \emph{ideal} control sequence $u_{[0,N-1]}^{\star}$ such that the following holds:
\begin{equation}\label{eq:ideal_input}
    \begin{aligned}
        x_{k+1}&=f(x_k,u_k^{\star}),\\
        y_{k}^d&=g(x_k),
    \end{aligned}~~~\qquad k=0,1,\ldots,N-1,
\end{equation}
i.e., the model-reference control objective is achieved. By relying on this assumption, we can formulate the direct \emph{in-context} control design problem as
\begin{equation}\label{eq:ideal_problem}
    \min_{\phi}~~\mathbb{E}_{p(\mathcal{D})}\left[\sum_{k=0}^{N-1}\|u_k^\star-\mathcal{C}_{\phi}(\mathcal{I}_k)\|_{2}^{2}\right]\!,
\end{equation}
where $p(\mathcal{D})$ indicates the distribution of the meta-dataset in \eqref{eq:meta_data}. Accordingly, the contextual controller is learned to be a direct mapping between current/past tracking errors and past inputs to the optimal control action needed to achieve the targeted reference model for \emph{any system} in $\mathcal{S}$. However, for the previous problem to be solved exactly and, hence, the contextual controller to \textquotedblleft understand\textquotedblright \ which input relates to a given context \eqref{eq:information_vector}, the controller's training should be carried out by using an infinite dimensional meta-dataset, i.e., $b=\infty$ in \eqref{eq:meta_data}. 

To overcome this first limitation, we rely on a finite (yet large) meta-dataset and approximate \eqref{eq:ideal_problem} as       
\begin{equation}\label{eq:true_problem1}
    \min_{\phi}~~ \frac{1}{b}\sum_{i=1}^{b}\sum_{k=0}^{N-1}\left\|u_k^{\star,(i)}-\mathcal{C}_{\phi}\left(\mathcal{I}_k^{(i)}\right)\right\|_{2}^{2},
\end{equation}
where $$\mathcal{I}_k^{(i)}=\left\{e_{\kappa}^{(i)},u_{\kappa-1}^{(i)}\right\}_{\kappa=0}^{k},$$ is the context associated with the $i$-th random draw from $p(\mathcal{D})$, $e_{\kappa}^{(i)}=r_{\kappa}^{(i)}-y_{\kappa}^{(i)}$ for $\kappa \in \{0,\ldots,k\}$, and $u_{[0,N-1]}^{\star,(i)}$ is the corresponding ideal input. 

\begin{remark}[Size of the meta-dataset]
Increasing the size $b$ of the meta-dataset is relatively simple if a simulator or a digital twin of the system class is available. Indeed, they can be queried unlimited times to gather data without interfering with the system's normal operation.    
\end{remark}

Although not exploiting an infinite dimensional dataset, the training procedure relying on \eqref{eq:true_problem1} still requires the ideal control sequences $u_{[0,N-1]}^{\star,(i)}$, $i=1,\ldots,b$. However, these sequences are onerous to retrieve even in simulation, as they would demand the design of a \textquotedblleft perfect\textquotedblright \ model reference controller for each of the $b$ random draws from $p(\mathcal{D})$. To overcome this additional limitation, we rely on two intuitions. When fed to the $i$-th system $S^{i}$
\begin{enumerate}
    \item the input sequence $u_{[0,N-1]}^{\star,(i)}$ results in the desired (and known) output $y_{[0,N-1]}^{d,(i)}$ to the reference $r_{[0,N-1]}^{(i)}$,
    \item the input generated by the in-context controller leads to the corresponding closed-loop response $y_{k,\phi}^{\mathrm{cl},(i)}$,
\end{enumerate}
for all $i=1,\ldots,b$. Accordingly, we propose to recast \eqref{eq:true_problem1} as  
\begin{equation}\label{eq:true_problem2}
    \min_{\phi}~~ \frac{1}{b}\sum_{i=1}^{b}\sum_{k=0}^{N-1}\left\|y_k^{d,(i)}-y_{k,\phi}^{\mathrm{cl},(i)}\right\|_{2}^{2},
\end{equation}
which ultimately represents our \emph{in-context control design} problem.
\begin{figure}[!tb]
  \begin{center}
     \includegraphics[scale=.75, trim=1cm 18.5cm 11cm 2cm]{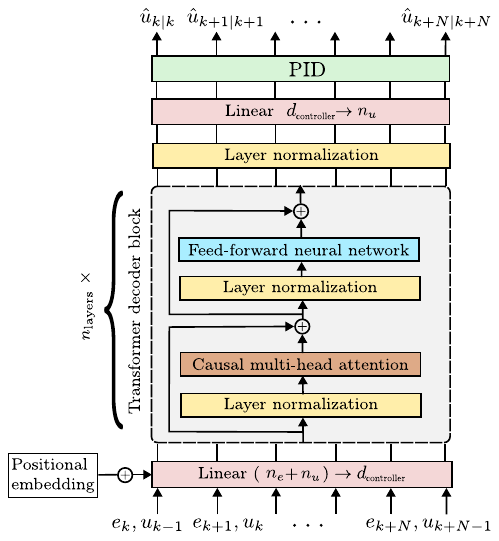}  \end{center}\vspace{-.6cm}
  \caption{GPT-like decoder-only Transformer for in-context control.}\label{fig:decoder}
\end{figure}
\section{Structuring and training a contextual controller}\label{sec:training}
With the learning problem \eqref{eq:true_problem2} at hand, we now provide additional insights into the structure of the in-context controller and its training procedure.
\subsection{On the structure of the in-context controller}
In the footsteps of \cite{forgione2023system,busetto2024context}, for the contextual controller, we use a \emph{decoder-only} Transformer architecture fed by current/past tracking errors and past inputs and outputting the control action to be fed to the controlled system. Nonetheless, different from the structures exploited for identification and estimation in \cite{forgione2023system} and \cite{busetto2024context}, respectively, we here tailor the \emph{decoder-only Transformer} to tackle a tracking task.
As schematized in \figurename{~\ref{fig:decoder}}, this customization is performed by introducing an output 
layer enforcing \emph{proportional}, \emph{integral}, and \emph{derivative} (PID) components in the control action. While this design choice minimally augments the number of parameters to be learned (as only three additional parameters are introduced), it promotes zero steady-state tracking error for step-like references and, based on our experience, it speeds up the training process.   
\subsection{How to train a contextual controller}
\begin{algorithm}[!tb]
\caption{Closed-loop simulation for contextual control design}\label{alg:cl_train}
\KwIn{$\bar{S} \sim p(\mathcal{S})$, $\bar{r}_{[0,N-1]} \sim p(\mathcal{R})$, $(\bar{x}_{0},\bar{u}_{-1}) \sim p(\mathcal{O})$, $\bar{\phi}$}
\KwOut{Residual sum of squares $\mathrm{RSS}(\bar{S},\bar{r}_{[0,N-1]},\bar{x}_{0};\bar{\phi})$}
$x_{0}^{M} \gets \bar{x}_{0}$;\\
\textbf{get} $y_{[0,N-1]}^{d}$ by feeding \eqref{eq:reference_model} with $\bar{r}_{[0,N-1]}$;\\
\For{$k = 0$ \KwTo $N-1$}{
    ~\textbf{get} the output $\bar{y}_{k,\bar{\phi}}^{\mathrm{cl}}$ of $\bar{S}$ given $\bar{x}_k$;\\
    $\bar{e}_k = \bar{r}_k - \bar{y}_{k,\bar{\phi}}^{\mathrm{cl}}$;\\
    $\bar{u}_{0:k} = \mathcal{C}_{\bar{\phi}}(\bar{e}_{0:k}, \bar{u}_{-1:k-1})$;\\
    \textbf{get} the updated state $\bar{x}_{k+1}$ of $\bar{S}$ by feeding it with $\bar{u}_k$; 
    }
$\mathrm{RSS}(\bar{S},\bar{r}_{[0,N-1]},\bar{x}_{0};\bar{\phi}) \gets \sum_{k=0}^{N-1}\|y_k^{d}-\bar{y}_{k,\bar{\phi}}^{\mathrm{cl}}\|_{2}^{2}$     
\end{algorithm}
Based on \eqref{eq:true_problem2}, our in-context control design formulation requires closing the loop (in simulation) on the systems drawn from $p(\mathcal{S})$ (see the procedure in Algorithm~\ref{alg:cl_train}). In turn, this implies that for end-to-end learning with back-propagation, the state-space representation of the systems $S^{(i)}$, $i=1,\ldots,b$, has to be differentiable for the required gradients to be computed. This feature does not allow us to use black-box simulators to construct the meta-dataset, differentiating the learning task from those for in-context identification and estimation. 

Considering a closed-loop learning procedure not only prevents us from using black-box simulators but, based on our experience, leads to poor results if the contextual controller is trained with the same strategy adopted in \cite{forgione2023system,busetto2024context}, i.e., feeding the Transformer by sampling from the overall distribution
\begin{equation}\label{eq:distribution}
    p(\mathcal{D})=p(\mathcal{S}) \times p(\mathcal{R}) \times p(\mathcal{O}).
\end{equation}
Instead of using this procedure, we employ \emph{curriculum learning} \cite{bengio2009curriculum} to increment the learning task's complexity gradually. As summarized in Algorithm~\ref{alg:training}, training begins with a first run in which the transformer is trained on a single system $\bar{S}$, reference sequence $\bar{r}_{[0,N-1]}$, and initial condition $\bar{x}_0$ drawn at random from the available distribution $p(\mathcal{D})$. This initial transformer is then refined by sampling over the class of systems via $p(\mathcal{S})$, yet keeping the reference and the initial condition fixed. Training is then performed again by sampling over both the class of systems and references and keeping the initial condition fixed, starting from the result of the previous iteration. A last training round is performed by sampling from the overall distribution $p(\mathcal{D})$, starting from the Transformer obtained at the previous iteration. 
\begin{algorithm}[!tb]
\caption{Curriculum learning for contextual controllers training}\label{alg:training}
\KwIn{$p(\mathcal{D})$ in \eqref{eq:distribution}, user-defined tolerance $\mathcal{L}_{\mathrm{min}}$, initial guess $\phi$, step size $\eta>0$, $\bar{S} \sim p(\mathcal{S})$, $\bar{r}_{[0,N-1]} \sim p(\mathcal{R})$, $(\bar{x}_{0},\bar{u}_{-1}) \sim p(\mathcal{O})$}
\KwOut{Optimal (in the mean square error sense) contextual controller $\mathcal{C}_{\phi^{\star}}(\cdot)$}
\For{$j=1$ \KwTo $4$}{
$\mathcal{L} \gets \infty$;\\
\While{$\mathcal{L}\geq \mathcal{L}_{\mathrm{min}}$}{
\uIf{$j=1$}{
    ~\textbf{compute} $\mathcal{L} \gets \mathrm{RSS}(\bar{S},\bar{r}_{[0,N-1]},\bar{x}_{0};\phi)$ by running Algorithm~\ref{alg:cl_train};
  }
\Else{
\For{$i=1$ \KwTo $b$}{
\uIf{$j=2$}{
~\textbf{get} $S^{(i)}\sim p(\mathcal{S})$;\\
\textbf{compute} $\mathcal{L}^{(i)} \gets \mathrm{RSS}(S^{(i)},\bar{r}_{[0,N-1]},\bar{x}_{0};\phi)$ by running Algorithm~\ref{alg:cl_train};
}
\uElseIf{$j=3$}{
~\textbf{get} $S^{(i)}\sim p(\mathcal{S})$ and $r_{[0,N-1]}^{(i)}\sim p(\mathcal{R})$;\\
\textbf{compute} $\mathcal{L}^{(i)} \gets \mathrm{RSS}(S^{(i)},r_{[0,N-1]}^{(i)},\bar{x}_{0};\phi)$ by running Algorithm~\ref{alg:cl_train};
}
\Else{
~\textbf{get} $S^{(i)}\sim p(\mathcal{S})$, $r_{[0,N-1]}^{(i)}\sim p(\mathcal{R})$ and $x_{0}^{(i)}\sim p(\mathcal{X}_0)$;\\
\textbf{compute} $\mathcal{L}^{(i)} \gets \mathrm{RSS}(S^{(i)},r_{[0,N-1]}^{(i)},x_{0}^{(i)};\phi)$ by running Algorithm~\ref{alg:cl_train};
}
}
~$\mathcal{L} \gets \frac{1}{b}\sum_{i=1}^{b} \mathcal{L}^{(i)}$;
}
$\phi \gets \phi-\eta \nabla \mathcal{L}$;
}
}
$\phi^{\star} \gets \phi$;
\end{algorithm}

\section{In-context control of an evaporation process: a numerical case study}\label{sec:example}
As a benchmark example to assess the effectiveness of the contextual controller, we consider the nonlinear evaporation process already used in \cite{busetto2024context}. In particular, the dynamics of the system under control\footnote{Dependencies on time $t \in \mathbb{R}$ are omitted for the sake of readability.} is given by (see \cite{amrit2013optimizing}): 
\begin{subequations}\label{eq:evaporator}
\begin{equation}
    \begin{aligned}
        M\dot{X}_2&=F_1X_{2}+F_2P_{2},\\
         C\dot{P}_2&=F_4+F_5,\\
         y&=X_{2},
    \end{aligned}
\end{equation}
where $X_2$~[\%] and $P_2$~[kPa] are the product composition and the operating pressure of the process at a given time, while $F_1$, $F_2$, $F_4$ and $F_5$ are nonlinear functions of the evaporator's state and inputs $P_{100}$~[kPa] and $F_{200}$~[kg/min], namely its stream pressure and the cooling water flow rate. Such nonlinear dependencies arise from the dynamics of individual components of the process, i.e.,
\begin{itemize}
    \item the liquid-energy balance of the process, dictated by
    \begin{equation}
        T_2=aP_{2}+bX_{2}+c,~~\quad T_3=dP_2+e,~~\quad F_{4}=\frac{Q_{100}-F_1C_p(T_2-T_1)}{\lambda};
    \end{equation}
    \item the heat steam jacket, governed by
    \begin{equation}
        T_{100}=\varphi P_{100}+\gamma,~~~ Q_{100}=UA_1(T_{100}-T_{2}),~~~UA_{1}=h(F_{1}+F_{3}),~~~ F_{100}=\frac{Q_{100}}{\lambda_{s}};
    \end{equation}
    \item the condenser, governed by
    \begin{equation}
        Q_{200}=\frac{UA_{2}(T_{3}-T_{200})}{1+UA_2/(2C_{p}F_{200})},~~\quad F_{5}=\frac{Q_{200}}{\lambda};
    \end{equation}
    \item the level controller, leading to
    \begin{equation}
        F_{2}=F_{1}-F_{4}.
    \end{equation}
\end{itemize}
\end{subequations}
While the system is in continuous time, the design will be performed in discrete time by collecting data at a sampling time $T_s$ of $1$ [s].

Our goal is to make any evaporation process match the desired behavior dictated by the reference model
\begin{equation}
\begin{aligned}
x_{k+1}^{M}&=0.4286x_{k}^{M}+0.7143 r_{k},\\
y_{k}^{d}&=0.5669x_{k}^{M}+0.2914 r_{k},
\end{aligned}
\end{equation}
by using the trained contextual controller.
\subsection{Meta-dataset construction}
\begin{table*}[!tb]
\caption{Nominal model parameters for the evaporation process.}
\label{tab:process_params}
    \centering
    \scalebox{.775}{\begin{tabular}{ccccccccccccccccccc}
        \hline
        $a$ & $b$ & $c$ & $d$ & $e$ & $\varphi$ & $\gamma$ & $h$ &
        $M$ & $C$ & $UA_2$ & $C_p$ & $\lambda$ & $\lambda_s$ & $F_1$ & $X_1$ &
        $F_3$ & $T_1$ & $T_{200}$\\
        \hline
        0.5616 & 0.3126 & 48.43 & 0.507 & 55 & 0.1538 & 55 & 0.16 &
        20 & 4 & 6.84 & 0.07 & 38.5 & 36.6 & 10 & 5 &
        50 & 40 & 25\\
        \hline
    \end{tabular}}
    \label{tab:nominal_param}
\end{table*}
\begin{figure}[!tb]
    \centering
    \includegraphics[scale=.6,trim=0cm .4cm 0cm 0cm,clip]{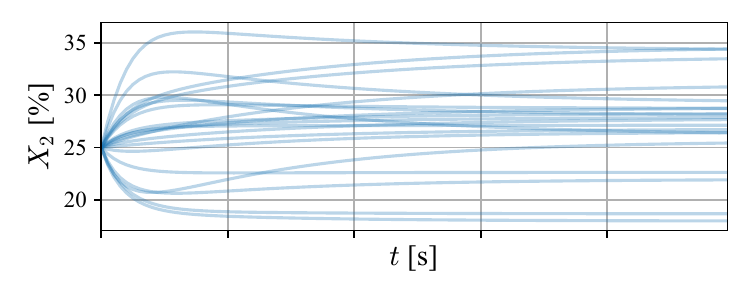}\vspace{-.2cm}
    \caption{Noiseless product composition over time of $20$ systems drawn at random from $\mathcal{S}$, starting from the same initial condition and subject to the same inputs.}
    \label{fig:meta_systems}
\end{figure}
The relations in \eqref{eq:evaporator} are characterized by 19 constants, whose nominal values are reported in \tablename{~\ref{tab:nominal_param}}. To construct the meta-dataset, these values are uniformly perturbed at random of up to 5~\% of their nominal values to construct the system's class $\mathcal{S}$ from which the instances of $\mathcal{D}$ are sampled at random. Note that, as shown in \figurename{~\ref{fig:meta_systems}}, such a perturbation leads to visibly different responses even when starting from the same initial condition and the same input\footnote{\label{footnote:4}The initial condition is $(X_2,P_2)=(25~\mathrm{[\%]},49.743~\mathrm{[kPa]})$, while the input considered is the steady-state input for the system with the nominal parameters reported in \tablename{~\ref{tab:nominal_param}}, namely $(P_{100},F_{200})=(191.713~\mathrm{[kPa]}, 215.888~\mathrm{[kg/min]})$ for all $t$.}.

Instead, the class of references we are interested in tracking is random steps of length $N=100$. As such, the references comprised in the meta-dataset are constructed accordingly by imposing
\begin{equation}\label{eq:reference_meta}
r_{[\tau,\tau+s^{(i)}]}^{(i)}=\rho_{\tau}^{(i)},~~~\tau=0,\ldots,N-s^{(i)}-1,
\end{equation}
with $\rho_{\tau}^{(i)}$ sampled uniformly at random from the interval $[20,25]$~[\%] and the step duration $s^{(i)}$ randomly sampled from a uniform distribution over the interval $[20,50]$~[s], for each $i=1,\ldots,b$. 

Lastly, for the sake of simplifying the learning task, we have assumed the initial conditions for the states and the inputs to be all the same and coincide with that of the nominal process (see footnote~\ref{footnote:4}). Hence, training is performed by not sampling over $p(\mathcal{O})$.
\subsection{Details on the contextual controller training strategy}
\begin{table}[!tb]
    \centering
    \caption{Hyper-parameters for the contextual controller structure: number of controller's parameters $n_{\mathrm{param}}$, number of layers of the decoder-only architecture $n_{\mathrm{layers}}$, number of heads $n_{\mathrm{heads}}$, length of the context $n_{\mathrm{ctx}}$, number of units in each layer $d_{\mathrm{controller}}$.}
    \label{tab:hyperparams}
    \scalebox{.8}{\begin{tabular}{ccccccc}
        \hline
        $n_{\mathrm{param}}$ & $n_{\mathrm{layers}}$ & $n_{\mathrm{heads}}$ & $n_{\mathrm{ctx}}$ & $d_{\mathrm{controller}}$ & $n_{\mathrm{itr}}$ & $b$
        \\
        \hline
        $10^{5}$ & 8 & 4 & 100 & 128 & 52 $\cdot 10^{3}$  & 1\\
        \hline
    \end{tabular}}
\end{table}
\begin{table}[!tb]
    \centering
    \caption{Number of iterations $n_{\mathrm{itr}}$, training time [h] and validation M-RMSE \eqref{eq:rmse_val} across training stages.}
    \label{tab:training_outcome}
    \scalebox{.8}{\begin{tabular}{lccc}
        \cline{2-4}
        & $n_{\mathrm{itr}}$ & training time [h] & M-RMSE 
        \\
        \hline
        Step 1 & 19780 & 36 & $7 \cdot 10^{-4}$\\
        \hline
        Step 2 & 6279 & 12 & $10^{-2}$\\
        \hline
        Step 3 & 26405 & 99 & $2 \cdot 10^{-2}$\\
        \hline
    \end{tabular}}
\end{table}
The previous design choice simplifies the curriculum learning routine summarized in Algorithm~\ref{alg:training}. In particular, it consists of Step 1 to Step 3 only, training first the contextual controller over the nominal system and a fixed reference, then considering new realizations of the evaporation process, and, lastly, training over new realizations of systems and references.

By setting the Transformer hyper-parameters reported in \tablename{~\ref{tab:hyperparams}} and training\footnote{The code is available at \url{https://github.com/buswayne/in-context-controller}} and starting from a random initial guess 
for its parameters
, the contextual controller is trained via the stochastic gradient descent strategy presented in  \cite{loshchilov2016sgdr} within the PyTorch framework \cite{paszke2019pytorch} on a Linux-based x86-64 server with 630 GB RAM and an NVIDIA RTX 3090 GPU\footnote{To avoid CPU induced bottlenecks, the data generation code is adapted to run entirely on the GPU.}. The number of iterations, training time, and the \emph{matching root mean square error} (M-RMSE), i.e., 
\begin{equation}\label{eq:rmse_val}
    \mathrm{M}\text{-}\mathrm{RMSE}=\sqrt{\frac{1}{N}\sum_{k=0}^{N-1}\left(y_{k}^{d,\mathrm{val}}-y_{k,\phi}^{\mathrm{cl},\mathrm{val}}\right)^{2}},
\end{equation}
obtained by validating the transformer on new instances of the system and tracking reference are reported in \tablename{~\ref{tab:training_outcome}}. Note that, as expected, the M-RMSE increases with the complexity of the learning task. Meanwhile, despite the number
of iterations overall performed for training resembles that needed in \cite{busetto2024context} to train an in-context estimator for the same example, the time required for the contextual controller to be trained is about $6$ days, due to the need for closed-loop simulations in training. Even if this increase in the training time is likely not an issue, as the computational effort has to be completely undertaken \emph{offline}, this paves the way to pursuing alternative (lighter) training schemes that we postpone to future works. 

\subsection{Performance analysis}
\begin{figure}[!tb]
    \centering
    \includegraphics[scale=0.6,trim=0cm 0cm 0cm 0cm,clip]{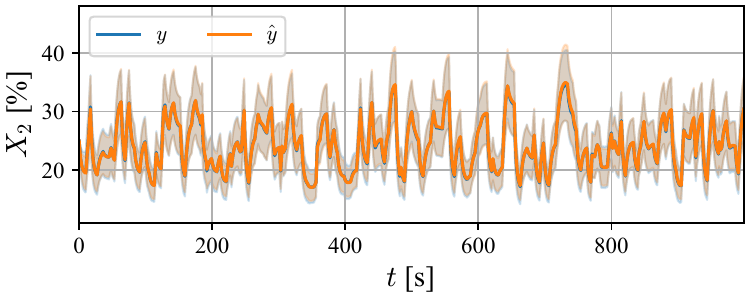}\vspace{-.2cm}
    \caption{Observed ($y$) \emph{vs} predicted ($\hat{y}$) composition with the identified grey-box models for the evaporation process: averages (lines) and standard deviations (shaded areas).}
    \label{fig:identification}
\end{figure}
We now consider a specific reference for the product composition and $20$ fresh evaporation processes, with the latter affected by white, zero-mean mutually independent process and measurement noise, both Gaussian distributed with covariance $0.1I$ (with $I$ being an identity matrix with proper dimensions) and variance $0.1$, respectively, for testing. 

As the contextual controller configures as a unique structure trained (offline) and then deployed to immediately regulate any new, unknown evaporation process toward achieving the given reference (closed-loop) behavior, it is difficult to find fair, available competitors already up for this task. We have thus decided to compare the contextual controller's performance with that achieved with two \emph{optimal control} (OC) schemes, i.e., by solving
\begin{equation}\label{eq:OC_testing}
    \begin{aligned}
        & \min_{P_{100},F_{200}}~\frac{1}{N_{H}}\sum_{k=0}^{N_{H}-1}\left(y_{k}^{d,\mathrm{test}}-y_{k}^{\mathrm{test}}\right)^{2}\\
        &\qquad \mbox{s.t.}~~\mathrm{eq. }\eqref{eq:evaporator} 
    \end{aligned}
\end{equation}
in a receding horizon fashion (with horizon $N_{H}$) over simulations of $100$ seconds, with the parameters dictating the (discretized) evaporator dynamics being either $(i)$ the \textquotedblleft true\textquotedblright \ ones of each evaporation process, leading to what we refer to as an \emph{oracle} scheme, or $(ii)$ \emph{identified} (id) in a gray-box fashion\footnote{The mean square error between the measured outputs and the one predicted with the identified model is optimized over a maximum of $100$ steps of stochastic gradient descent \cite{loshchilov2016sgdr} always starting from the nominal parameters in \tablename{~\ref{tab:nominal_param}} for all the testing systems and terminating the identification procedure whenever the objective is below $10^{-4}$.} from a set of open-loop $1000$ input/output data\footnote{The systems are excited with random sequences of steps sampled uniformly within the interval $[150,250]$, with random length uniformly distributed in $[1,10]$ [s].}. Such a procedure leads to the difference between true and estimated outputs on a fresh test set of open-loop input/output data shown in \figurename{~\ref{fig:identification}}, corresponding to a root mean square output prediction error of $0.7479$. This last choice of a competitor has been made since it mimics the simplest-to-deploy (model-based) solution for model-reference control of a nonlinear system, while we demand comparisons with alternative competitors to future works.
\begin{figure}[!tb]
    \centering
    \includegraphics[scale=.6,trim=0cm .2cm 0cm .1cm,clip]{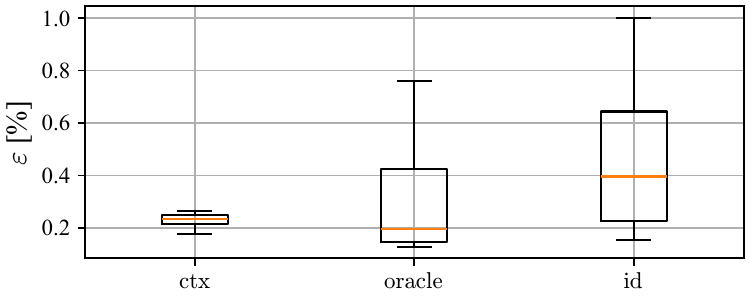}\vspace{-.2cm}
    \caption{Boxplots of the matching errors achieved over the $20$ performed closed-loop tests with the contextual (ctx), oracle, and identification-based (id) controllers.}
    \label{fig:boxplot}
\end{figure}
\begin{figure}[!tb]
    \centering
    \begin{tabular}{cc}
    \includegraphics[scale=.55,trim=0cm 0cm 0cm 0cm,clip]{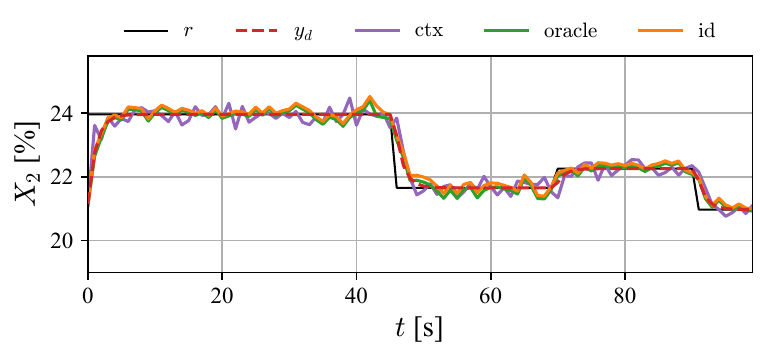} &
    \includegraphics[scale=.55,trim=0cm 0cm 0cm 0cm,clip]{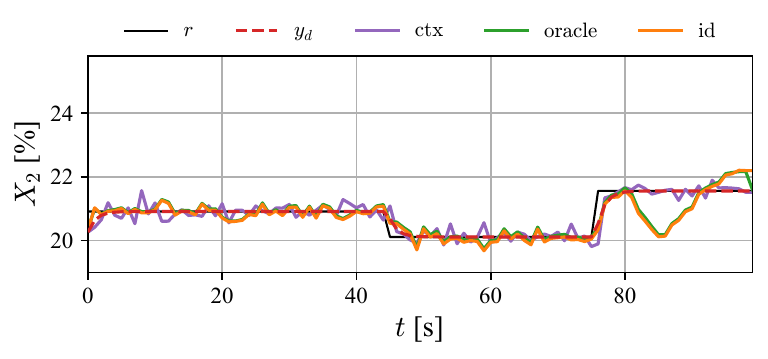}\vspace{-.2cm}\\
    \multicolumn{2}{c}{\includegraphics[scale=.55,trim=0cm 0.1cm 0cm 0cm,clip]{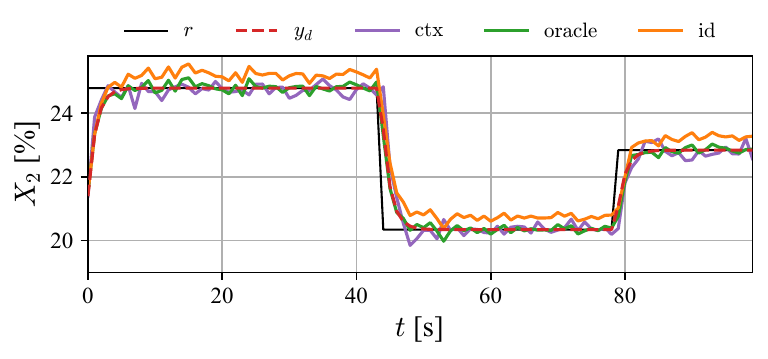}} 
    \end{tabular}\vspace{-.2cm}
    \caption{Closed-loop responses attained by the contextual (ctx), oracle, and identification-based (id) controllers \emph{vs} reference to be tracked and desired behavior over $3$ (randomly drawn) test out of the $20$ we have performed.}
    \label{fig:expl_boxplot}
\end{figure}

We first set the horizon for the OC policies to $N_{H}=5$, obtaining (over the $20$ test evaporation processes and references) the matching errors
\begin{equation}
    \varepsilon=\left\|y_{[0,99]}^{d,\mathrm{test}}-y_{[0,99]}^{\mathrm{cl},\mathrm{test}}\right\|_{2},~~~[\%]
\end{equation}
summarized in \figurename{~\ref{fig:boxplot}}. These results confirm that the contextual controller can generalize to different instances of the evaporation processes and the reference product compositions, outperforming the considered model-based competitors. Such superior performance can be explained by looking at the results shown in \figurename{~\ref{fig:expl_boxplot}}, where we report the responses attained with the contextual controller and the OC schemes for 4 (randomly selected) closed-loop tests. While the responses attained by the \emph{oracle} and the contextual controller are generally similar, the ones achieved by relying on identified models often deteriorate (see the second and third panels of \figurename{~\ref{fig:expl_boxplot}}), likely due to the accuracy of such models. Moreover, both the OC schemes are subject to isolated cases in which the problem is not feasible (see the second panel of \figurename{~\ref{fig:expl_boxplot}}), leading to an overall worst matching.   

\begin{figure}[!tb]
    \centering
    \begin{tabular}{cc}
    \subfigure[Contextual controller]{\includegraphics[scale=.6,trim=0cm 0.cm 0cm 0cm,clip]{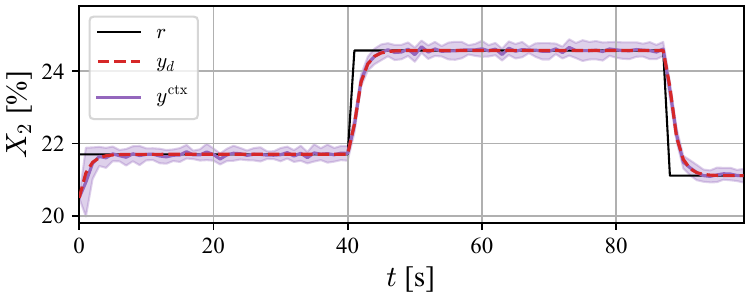}\label{fig:context_performance}} & \subfigure[Oracle controller]{\includegraphics[scale=.6,trim=0cm 0cm 0cm 0cm,clip]{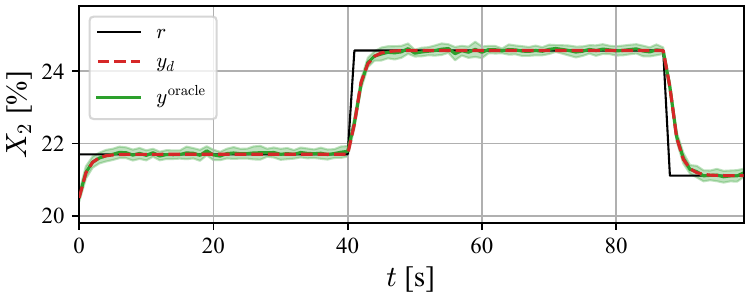}\label{fig:nmpc_oracle_performance}}\\
    \multicolumn{2}{c}{\subfigure[Identification-based controller]{\includegraphics[scale=.6,trim=0cm 0cm 0cm 0cm,clip]{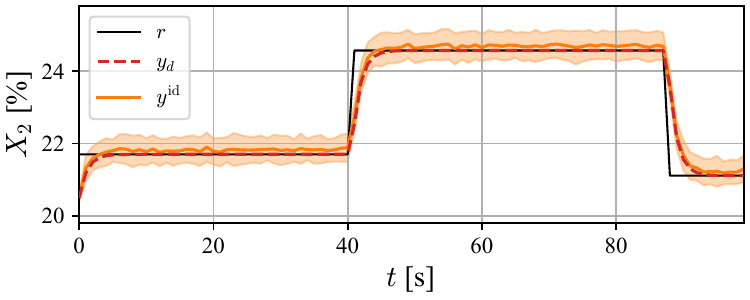}\label{fig:nmpc_identified_performance}}} 
    \end{tabular}\vspace{-.1cm}
    \caption{Average (colored lines) and standard deviation (shaded areas) of the closed-loop responses attained by the contextual, oracle, and identification-based controllers \emph{vs} reference to be tracked and desired behavior over the $20$ tests performed.}
    \label{fig:CL_comparisons}
\end{figure}
By increasing the horizon $N_{H}$ of the OC schemes to $10$, these last instances can be removed, leading to the closed-loop compositions shown in \figurename{~\ref{fig:CL_comparisons}}. Clearly (and as one would have expected), the oracles now achieve better performance with respect to the contextual controller. At the same time, this result highlights the reliance of the oracles' performance on a careful hyper-parameter selection, which is instead not required (after training) when deploying the contextual controller. Moreover, the latter still outperforms the controllers relying on the identified models except for (approximately) the first $10$ [s] of closed-loop testing (see the variability of the closed-loop response in \figurename{~\ref{fig:CL_comparisons}}). This last result can nevertheless be expected as the contextual controller has to initially understand the \textquotedblleft context\textquotedblright \ from the incoming closed-loop data to discriminate the best control action to feed each specific instance of the evaporation process.

For completeness, it is worth pointing out that the average inference time of the contextual controller is about 2.8 [ms], while our greedy approach to solving \eqref{eq:OC_testing} takes 1 to 10 [s] at each simulation step\footnote{These computational times can be further improved by refining the OC implementation.}. 

\section{Conclusions and future works}\label{sec:conclusions}
After the promising results achieved in system identification (\cite{forgione2023system}) and estimation (\cite{busetto2024context}), this work extends the exploration of the in-context learning framework for control by presenting the \emph{contextual controller}, a unique mapping from tracking errors/inputs to new control actions to regulate \emph{all} systems within a class toward a desired closed-loop behavior. The results achieved on a benchmark numerical example highlight the capability of this solution for immediate regulation of new systems within the class without fine-tuning, often achieving superior performance with respect to model-based receding horizons control policies.  

Future research will involve guaranteeing stability and performance by design for loops closed with the contextual controller, testing the current architecture on real-world applications, where the idealities of \emph{in-silico} examples are removed, pursuing architectural changes to reduce the time required for the contextual controller to understand the context, and exploring alternative training strategies to ease the (currently computationally demanding) training procedure.


\bibliography{main}

\begin{thebibliography}{23}
\providecommand{\natexlab}[1]{#1}
\providecommand{\url}[1]{\texttt{#1}}
\expandafter\ifx\csname urlstyle\endcsname\relax
  \providecommand{\doi}[1]{doi: #1}\else
  \providecommand{\doi}{doi: \begingroup \urlstyle{rm}\Url}\fi

\bibitem[Aky{\"u}rek et~al.(2024)Aky{\"u}rek, Wang, Kim, and Andreas]{akyurek2024context}
Ekin Aky{\"u}rek, Bailin Wang, Yoon Kim, and Jacob Andreas.
\newblock In-context language learning: Architectures and algorithms.
\newblock \emph{arXiv preprint arXiv:2401.12973}, 2024.

\bibitem[Amrit et~al.(2013)Amrit, Rawlings, and Biegler]{amrit2013optimizing}
Rishi Amrit, James~B Rawlings, and Lorenz~T Biegler.
\newblock Optimizing process economics online using model predictive control.
\newblock \emph{Computers \& Chemical Engineering}, 58:\penalty0 334--343, 2013.

\bibitem[Bansal et~al.(2022)Bansal, Alzubi, Wang, Lee, and McCallum]{bansal2022meta}
Trapit Bansal, Salaheddin Alzubi, Tong Wang, Jay-Yoon Lee, and Andrew McCallum.
\newblock Meta-adapters: Parameter efficient few-shot fine-tuning through meta-learning.
\newblock In \emph{International Conference on Automated Machine Learning}, pages 19--1. PMLR, 2022.

\bibitem[Bengio et~al.(2009)Bengio, Louradour, Collobert, and Weston]{bengio2009curriculum}
Yoshua Bengio, J{\'e}r{\^o}me Louradour, Ronan Collobert, and Jason Weston.
\newblock Curriculum learning.
\newblock In \emph{Proceedings of the 26th annual international conference on machine learning}, pages 41--48, 2009.

\bibitem[Brown et~al.(2020)Brown, Mann, Ryder, Subbiah, Kaplan, Dhariwal, Neelakantan, Shyam, Sastry, Askell, Agarwal, Herbert-Voss, Krueger, Henighan, Child, Ramesh, Ziegler, Wu, Winter, Hesse, Chen, Sigler, Litwin, Gray, Chess, Clark, Berner, McCandlish, Radford, Sutskever, and Amodei]{brown2020language}
Tom Brown, Benjamin Mann, Nick Ryder, Melanie Subbiah, Jared~D Kaplan, Prafulla Dhariwal, Arvind Neelakantan, Pranav Shyam, Girish Sastry, Amanda Askell, Sandhini Agarwal, Ariel Herbert-Voss, Gretchen Krueger, Tom Henighan, Rewon Child, Aditya Ramesh, Daniel Ziegler, Jeffrey Wu, Clemens Winter, Chris Hesse, Mark Chen, Eric Sigler, Mateusz Litwin, Scott Gray, Benjamin Chess, Jack Clark, Christopher Berner, Sam McCandlish, Alec Radford, Ilya Sutskever, and Dario Amodei.
\newblock Language models are few-shot learners.
\newblock In \emph{Advances in Neural Information Processing Systems}, pages 1877--1901. Curran Associates, Inc., 2020.

\bibitem[Busetto et~al.(2024)Busetto, Breschi, Forgione, Piga, and Formentin]{busetto2024context}
Riccardo Busetto, Valentina Breschi, Marco Forgione, Dario Piga, and Simone Formentin.
\newblock In-context learning of state estimators.
\newblock \emph{IFAC-PapersOnLine}, 58\penalty0 (15):\penalty0 145--150, 2024.

\bibitem[Campi and Savaresi(2006)]{campi2006direct}
Marco~C Campi and Sergio~M Savaresi.
\newblock Direct nonlinear control design: The virtual reference feedback tuning ({VRFT}) approach.
\newblock \emph{IEEE Transactions on Automatic Control}, 51\penalty0 (1):\penalty0 14--27, 2006.

\bibitem[Dong et~al.(2022)Dong, Li, Dai, Zheng, Wu, Chang, Sun, Xu, and Sui]{dong2022survey}
Qingxiu Dong, Lei Li, Damai Dai, Ce~Zheng, Zhiyong Wu, Baobao Chang, Xu~Sun, Jingjing Xu, and Zhifang Sui.
\newblock A survey on in-context learning.
\newblock \emph{arXiv preprint arXiv:2301.00234}, 2022.

\bibitem[Farahani et~al.(2021)Farahani, Voghoei, Rasheed, and Arabnia]{farahani2021brief}
Abolfazl Farahani, Sahar Voghoei, Khaled Rasheed, and Hamid~R Arabnia.
\newblock A brief review of domain adaptation.
\newblock \emph{Advances in data science and information engineering: proceedings from ICDATA 2020 and IKE 2020}, pages 877--894, 2021.

\bibitem[Finn et~al.(2017)Finn, Abbeel, and Levine]{finn2017model}
Chelsea Finn, Pieter Abbeel, and Sergey Levine.
\newblock Model-agnostic meta-learning for fast adaptation of deep networks.
\newblock In \emph{International conference on machine learning}, pages 1126--1135. PMLR, 2017.

\bibitem[Forgione et~al.(2023)Forgione, Pura, and Piga]{forgione2023system}
Marco Forgione, Filippo Pura, and Dario Piga.
\newblock From system models to class models: An in-context learning paradigm.
\newblock \emph{IEEE Control Systems Letters}, 2023.

\bibitem[Goel and Bartlett(2024)]{goel2024can}
Gautam Goel and Peter Bartlett.
\newblock Can a transformer represent a kalman filter?
\newblock In \emph{6th Annual Learning for Dynamics \& Control Conference}, pages 1502--1512. PMLR, 2024.

\bibitem[Hou and Wang(2013)]{hou2013model}
Zhong-Sheng Hou and Zhuo Wang.
\newblock From model-based control to data-driven control: Survey, classification and perspective.
\newblock \emph{Information Sciences}, 235:\penalty0 3--35, 2013.

\bibitem[Khoee et~al.(2024)Khoee, Yu, and Feldt]{khoee2024domain}
Arsham~Gholamzadeh Khoee, Yinan Yu, and Robert Feldt.
\newblock Domain generalization through meta-learning: a survey.
\newblock \emph{Artificial Intelligence Review}, 57\penalty0 (10):\penalty0 285, 2024.

\bibitem[Li et~al.(2018)Li, Yang, Song, and Hospedales]{li2018learning}
Da~Li, Yongxin Yang, Yi-Zhe Song, and Timothy Hospedales.
\newblock Learning to generalize: Meta-learning for domain generalization, 2018.

\bibitem[Ljung(1999)]{ljung1998system}
Lennart Ljung.
\newblock \emph{System identification (2nd ed.): theory for the user}.
\newblock Prentice Hall PTR, USA, 1999.
\newblock ISBN 0136566952.

\bibitem[Loshchilov and Hutter(2016)]{loshchilov2016sgdr}
Ilya Loshchilov and Frank Hutter.
\newblock Sgdr: Stochastic gradient descent with warm restarts.
\newblock \emph{arXiv preprint arXiv:1608.03983}, 2016.

\bibitem[Narvekar et~al.(2020)Narvekar, Peng, Leonetti, Sinapov, Taylor, and Stone]{narvekar2020curriculum}
Sanmit Narvekar, Bei Peng, Matteo Leonetti, Jivko Sinapov, Matthew~E Taylor, and Peter Stone.
\newblock Curriculum learning for reinforcement learning domains: A framework and survey.
\newblock \emph{Journal of Machine Learning Research}, 21\penalty0 (181):\penalty0 1--50, 2020.

\bibitem[Paszke et~al.(2019)Paszke, Gross, Massa, Lerer, Bradbury, Chanan, Killeen, Lin, Gimelshein, Antiga, et~al.]{paszke2019pytorch}
Adam Paszke, Sam Gross, Francisco Massa, Adam Lerer, James Bradbury, Gregory Chanan, Trevor Killeen, Zeming Lin, Natalia Gimelshein, Luca Antiga, et~al.
\newblock Pytorch: An imperative style, high-performance deep learning library.
\newblock \emph{Advances in neural information processing systems}, 32, 2019.

\bibitem[Vaswani(2017)]{vaswani2017attention}
A~Vaswani.
\newblock Attention is all you need.
\newblock \emph{Advances in Neural Information Processing Systems}, 2017.

\bibitem[Wang and Zhu(2023)]{wang2023context}
Xuan Wang and Zhigang Zhu.
\newblock Context understanding in computer vision: A survey.
\newblock \emph{Computer Vision and Image Understanding}, 229:\penalty0 103646, 2023.

\bibitem[Xu et~al.(2020)Xu, Zhang, Mao, Wang, Xie, and Zhang]{xu2020curriculum}
Benfeng Xu, Licheng Zhang, Zhendong Mao, Quan Wang, Hongtao Xie, and Yongdong Zhang.
\newblock Curriculum learning for natural language understanding.
\newblock In \emph{Proceedings of the 58th Annual Meeting of the Association for Computational Linguistics}, pages 6095--6104, 2020.

\bibitem[Zhu et~al.(2024)Zhu, Cano, Bermudez, and Drozdzal]{zhu2024incoro}
Jiaqiang~Ye Zhu, Carla~Gomez Cano, David~Vazquez Bermudez, and Michal Drozdzal.
\newblock Incoro: In-context learning for robotics control with feedback loops.
\newblock \emph{arXiv preprint arXiv:2402.05188}, 2024.

\end{thebibliography}

\end{document}